\documentclass[aps,prb, singlecolumn,superscriptaddress, longbibliography, nobibnotes]{revtex4-2}
\usepackage{graphicx}
\usepackage{color}
\usepackage{float}
\usepackage{amsmath}

\usepackage[breaklinks]{hyperref}
\hypersetup{
	pdfnewwindow=true,      
    colorlinks=true,       
    linkcolor=blue,          
    citecolor=blue,        
    filecolor=blue,      
    urlcolor=blue           
}

\begin{document}
\title{Nano-optics of transition metal dichalcogenides and their van der Waals heterostructures with electron spectroscopies}

\author{Steffi Y. Woo}
\email[Authors to whom any correspondence should be addressed: ]{woosy@ornl.gov}
\affiliation{Universit\'e Paris-Saclay, CNRS, Laboratoire de Physique des Solides, 91405, Orsay, France}
\affiliation{Center for Nanophase Materials Sciences, Oak Ridge National Laboratory, Oak Ridge, TN, 37381, U.S.A.}

\author{Luiz~H.~G.~Tizei}
\email[Authors to whom any correspondence should be addressed: ]{luiz.tizei@cnrs.fr}
\affiliation{Universit\'e Paris-Saclay, CNRS, Laboratoire de Physique des Solides, 91405, Orsay, France}
\date{\today}

\maketitle

\textbf{Abstract: }The outstanding properties of transition metal dichalcogenide (TMD) monolayers and their van der Waals (vdW) heterostructures, arising from their structure and the modified electron-hole Coulomb interaction in two-dimension, make them promising candidates for potential electro-optical devices. However, the production of reproducible devices remains challenging, partly due to variability at the nanometer to atomic scales. Thus, access to chemical, structural, and optical characterization at these lengthscales is essential. While electron microscopy and spectroscopy can provide chemical and structural data, accessing the optical response at the nanoscale through electron spectroscopies has been hindered until recently. This review focuses on the application of two electron spectroscopies in scanning (transmission) electron microscopes, namely cathodoluminescence and electron energy-loss spectroscopy, to study the nano-optics of TMD atomic layers and their vdW heterostructures. How technological advancements that can improve these spectroscopies, many of which are already underway, will make them ideal for studying the physics of vdW heterostructures at the nanoscale will also be discussed.

\section{Introduction}

Semiconducting transition metal dichalcogenide (TMD) monolayers and their van der Waals (vdW) heterostructures have garnered significant interest following the observation of bright photoluminescence (PL) from monolayers even at room temperature given their direct bandgap \cite{Mak2010}. These materials in the \textit{2H} phase with the form of $MX_2$ (where $M$ = Mo, W; $X$ = S, Se, Te) exhibit strong spin-orbit interaction from the transition metal, resulting in spin-valley states \cite{Xiao2012} at two triplets of inequivalent $K$ and  $K^\prime$ points. These anti-symmetric spin states lead to valley polarization, which can be selectively excited optically with circularly polarized photons \cite{Mak2012}. In addition, the exciton binding energy is large due to the increased Coulomb interaction in two-dimension (2D) and the weak electromagnetic screening at the 2D limit \cite{Chernikov2014,Thygesen2017,Wang2018}. Light-matter interactions are thus dominated by excitonic effects that result in strong absorption and emission resonances, making TMDs promising candidates for potential electro-optical devices. Furthermore, single-photon sources have been identified in these materials \cite{Tonndorf2015,Klein2019,Parto2021,Kim2023}, which can currently be fabricated with sub-100 nm precision using ion and electron beams. The nature of these sources remain highly debated, with the most likely hypothesis being impurity-bound defect states, local strain, or a combination of the two \cite{Linhart2019}. The atomic structures of these defect centers have also not yet been identified.

Reproducible production of clean devices is not yet a reality due to challenges in producing large (greater than 100 $\mu m^2$) clean monolayers and fabricating them into devices. Consequently, there is a need for techniques that allow for the structural and chemical characterization of vdW heterostructures, along with their optical properties, with nanometer spatial resolution.
Remarkable progress has been made with scanning probe techniques, such as tip-enhanced photoluminescence \cite{Lee2020}, and scanning tunneling microscope-induced luminescence (STML) \cite{Pommier2019, Roman2022, Lopez2023, Wang2023}. The demonstration of STML of defects in WS$_2$ \cite{Schuler2020} has sparked significant interest. One of the major advantages of STML is its ability to provide a direct connection to atomically-resolved imaging of surfaces and scanning tunneling spectroscopy (STS), which gives access to the material's single-particle energy band structure and bandgap \cite{Ugeda2014, Wang2018}.

This review will instead focus on two electron spectroscopies accessible in scanning and scanning transmission electron microscopes that provide nanometer-scale optical information on TMD atomic layers when performed with nanometer-wide electron beams: electron energy-loss spectroscopy (EELS) and cathodoluminescence (CL). These techniques have recently seen success in the field of nano-optics \cite{Polman2019}. For completeness, electron energy gain spectroscopy (EEGS) and temporal coincidence EELS and CL spectroscopy will also be mentioned, even though these techniques have not yet been exploited for TMD vdW heterostructures.

The text is organized as follows. Section \ref{sec::methods} provides background on the instrumentation and experimental techniques. Section \ref{sec::init_eels} discusses initial EELS experiments and explains why exciton linewidths in these experiments were significantly larger than in optical measurements. Section \ref{sec::bilayers} explores how the moir\'e angle in bilayer systems modifies the energy of excitonic transitions measured in EELS. The subsequent two sections describe how h-BN encapsulation, which is used to achieve CL light emission from TMD vdW heterostructures (Section \ref{sec::cl}), has led to a significant improvement in excitonic energy widths in EELS experiments (Section \ref{sec::current_eels}). Section \ref{sec::TMDcoupling} describes how EELS and CL can contribute to the understanding of exciton coupling in TMD heterostructures. Finally, Section \ref{sec::perspectives} considers how recent advancements in electron spectroscopy can impact the future research direction in TMDs.

Please note that this review does not delve into the details of imaging and diffraction experiments in these materials, unless they are related to the spectroscopy experiments described herein. Additionally, this review does not cover spectroscopic experiments for elemental mapping using EELS and energy-dispersed X-ray spectroscopy or probing the local density of unoccupied electron states (via EELS). Other comprehensive reviews that cover these topics on TMDs include ref. \cite{Wang2018EMreview,Luo2024}. For a more in-depth overview of TMD properties in general, including crystal structure, band structure, excitonic effects, etc., the reader can refer to detailed reviews in the literature that cover these topics, such as ref. \cite{Wang2018,Shree2021,Arora2021,Huang2022review,Glazov2024}.


\section{Instrumentation and Experimental Methods}
\label{sec::methods}

In this section, the experimental techniques are briefly described. Initially, Section \ref{sec::methods::SEM} provides an overview of scanning electron microscopy (SEM) and scanning transmission electron microscopy (STEM) with a focus on the instrumentation. This is followed by four sections, from \ref{sec::methods::eels} to \ref{sec::methods::cle}, which aim to provide an overview of electron energy-loss spectroscopy (EELS), cathodoluminescence (CL), electron energy gain spectroscopy (EEGS), and cathodoluminescence excitation spectroscopy (CLE) for materials science. Finally, Section \ref{sec::methods::samples} describes various methods for preparing TMD samples that are compatible with TEM.

\subsection{Scanning (transmission) electron microscopes}
\label{sec::methods::SEM}

Electron microscopes were envisioned and developed in the 1930s, based on the concept of using the shorter wavelength of electrons (3.7 pm for electrons with 100 keV kinetic energy) to overcome the diffraction limit imposed on light microscopy \cite{Ruska1987}. Initially, electron microscopes were designed for experiments with wide electron beams (on the micrometer scale) for “projection” experiments, where the electron beam is transmitted through the sample. In these transmission electron microscopes (TEM), the contrast in images arises due to phase changes imposed on the electron beam during propagation through the material. Experiments with this type of microscope are not discussed in this review. Therefore, interested readers are referred to seminal books on the subject and the references therein \cite{Kohl2008,Carter2016}.

Alternatively, electron microscopes can be operated with a focused electron beam that is raster-scanned over a sample. These scanning electron microscopes are divided, somewhat artificially, into two categories: scanning electron microscopes (SEM) and scanning transmission electron microscopes (STEM). Historically, these two were differentiated by the kinetic energy of the electron beam, with SEMs operating below 30 keV and STEMs above 100 keV. This limited SEMs to the study of the surfaces of bulk samples (volumes a few micrometers deep) and STEMs to thin samples (below a few hundred nanometers). Another historical difference was the accessibility of atomic resolution for STEMs operating above 200 keV. However, the development of aberration correctors has paved the way for atomically resolved STEM at 100 keV and below \cite{Krivanek2010}, blurring the boundary between SEM and STEM.

\begin{figure} [H]
\centering{\includegraphics[width=0.5\textwidth]{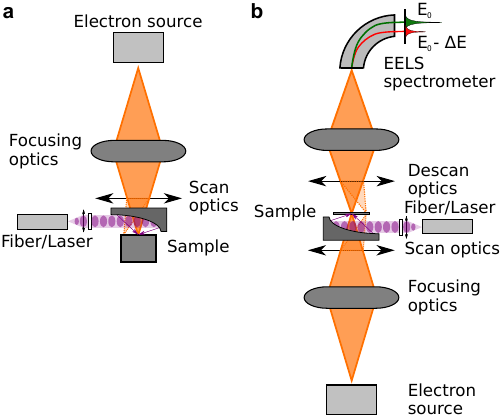}}
\caption{\textbf{a} Sketch of an SEM which includes an electron source, electron focusing optics (mostly purely magnetic), scanning optics (sequence of dipoles), a sample, and a light injection/collection system (CL). \textbf{b} Sketch of a STEM which, in addition to the elements in an SEM, contains electron beam de-scanning optics (EELS, diffraction), possibly a projection system, and an EELS spectrometer. The typical electron kinetic energy for SEMs is in the range of 1--30 keV, while for STEMs it is in the range of 30--300 keV. Samples in both setups can be cooled down to liquid-helium temperature, but this option is not standard for STEMs due to large mechanical vibrations and the short operation time of currently available helium-cooled sample holders.
\label{fig::SEM-STEM}}
\end{figure}

The principles of the two types of microscopes are very similar (Fig. \ref{fig::SEM-STEM}). An electron beam is generated by an electron source \cite{Kohl2008}. A cold field emitter (field-emission gun, FEG) with high brightness is needed for experiments requiring the highest spatial and spectral resolutions. The electron beam current, lateral size, and convergence angle are set by magnetic lenses in the condenser system (electrostatic lenses can be used for low-energy electrons, but for fast electrons, magnetic lenses are used due to the scaling as $\Vec{v}\times\Vec{B}$ of the Lorentz force). The electron beam is focused onto the sample using an objective lens, which is also a magnetic lens with cylindrical symmetry. Magnetic or electric multipoles located before the sample are used to correct for the geometric aberrations. After the sample, a system of projector lenses can be used in a STEM to adjust the camera length onto various detectors providing an array of signals that differ due to the electron scattering process. The various electron spectroscopy techniques will be discussed separately in subsequent sections, but a brief overview of their instrumentation will first be described in the following as there are common elements. A magnetic prism is used to disperse electrons that have transmitted past the sample in energy (for EELS and EEGS). A light injector/collector system (such as a flat or parabolic mirror) is used for CL and EEGS. Finally, electrons are detected (for EELS) using detectors coupled to scintillators, such as a charge-coupled device (CCD) or complementary metal-oxide-semiconductor (CMOS) camera, or direct electron detectors.

For optical spectroscopy, sample cooling is generally recommended to minimize thermal broadening. Some of the experiments described in the subsequent sections were performed with samples cooled by liquid nitrogen or helium, while others were performed at room temperature. Regardless, despite the existence of helium-cooled sample holders for electron microscopes, very few experiments with TMD vdW heterostructures have been conducted so far and only in an SEM \cite{Zheng2017,Francaviglia2022, Fiedler2023}.

\subsection{Electron energy-loss spectroscopy - EELS}
\label{sec::methods::eels}
\begin{figure} [H]
\centering{\includegraphics[width=0.4\textwidth]{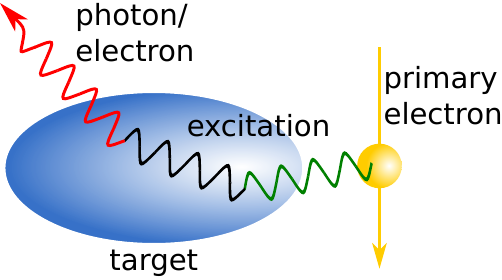}}
\caption{Sketch of inelastic scattering on a target. A primary electron can exchange energy with a target through its electromagnetic field at a finite distance, an effect usually referred to as delocalization \cite{Muller1995}. The energy lost creates excitations in the target, which can propagate. These excitations eventually decay, leading to the emission of phonons in the lattice and/or free photons and electrons. Spectroscopy of the energy lost (EELS) and the emitted photons (CL) helps us understand the optical properties of the target.
\label{fig::Scattering_Sketch}}
\end{figure}

When a fast electron is transmitted through a sample (Fig. \ref{fig::Scattering_Sketch}), it has a certain probability of losing energy through inelastic scattering \cite{Egerton2011}. The number of inelastic scattering events follows a Poisson distribution with parameter $\lambda_i(K)$, which represents the mean free path of electrons with kinetic energy $K$. The energy lost by the electron to excitations in the sample at each scattering event is not fixed, leading to a broadband spectrum \cite{Egerton2011}. This is an important benefit as well as a limitation for electron spectroscopies: it enables the measurement of properties in energy ranges that are hardly accessible to photon techniques (for example, the vacuum ultraviolet (VUV) range, between the UV and the soft X-ray energies), but it prevents resonant or energy-selective excitations (which are possible using photons).

Given the broadband energy range accessible, phonons, excitons, interband transitions, bulk and surface plasmons, and core electrons can all be excited \cite{Egerton2011}. The loss probability for the electron (EELS spectrum) in a bulk sample is proportional to $Im[-1/\varepsilon(\omega)]$ \cite{Egerton2011, Abajo2010}, where $\varepsilon(\omega)$ is the dielectric function at frequency $\omega$. However, for strictly 2D layers, because of the lack of electromagnetic screening, the loss probability is proportional to $Im[\varepsilon(q,\omega)]$ \cite{Kociak2001} making direct comparisons to optical extinction justified. For bulk materials $\varepsilon(q,\omega)$ can be retrieved using the Kramers-Kronig transformation \cite{Egerton2011}, even though this procedure is usually not trivial due to multiple scattering and boundary condition issues.
EELS spectra can also be measured as a function of momentum transfer, $q$. The measurement of excitation dispersion curves as a function of momentum transfer has been used to identify their nature, as demonstrated for the bulk plasmon in aluminum films \cite{Batson1983}.

Core-electron excitation allows chemical information to be gathered from EELS, including the quantification of the elemental species ratio, if their electron excitation cross-sections are known \cite{Egerton2011}. Also, the near-edge fine structure of absorption edges carries information about the local symmetry-projected density of unoccupied electronic states and hence about the atom's chemical structure and bonding \cite{Egerton2011}.

Practically, electron energy spectra in electron microscopes are measured using in-column omega filters or magnetic prisms at the end of the microscope column \cite{Egerton2011}. EELS spectrometers were previously not available for SEMs in commercial instruments, a situation that has recently changed \cite{Sunaoshi2016}.These magnetic elements disperse the electron beam in energy: electrons of different kinetic energy have different trajectories, meaning that the position after propagation encodes energy. A detector array (CCD/CMOS with a scintillator or a direct electron detector) can be used to measure histograms of the number of electrons per energy bin. One crucial point, which is often ignored, is the accuracy of the energy scale calibration. This can be obtained using standards or a potential change in the spectrometer’s drift tube, which leads to a known energy change. The first method depends on the availability of known and precise standards, which are still lacking for EELS in the visible range. The second method has a limited precision of $\sim$1\% at best. More technical details can be found elsewhere, including discussions about electron detectors in ref. \cite{Egerton2011}.

In EELS, two important limitations exist because of the energy spread of the primary electron beam. The peak associated with these electrons, referred to in the EELS community as zero-loss peak (ZLP), has an intrinsic linewidth and a tail towards larger energy losses (lower electron kinetic energy) for a cold FEG. The first property limits the minimum separation between two spectral features to be distinguished (referred to as energy resolution). The second one leads to a background intensity, limiting the detectability of weak signals, such as phonons \cite{Tian2021} and excitons in TMDs. An electron monochromator facilitates the decrease of both the linewidth and the tail of the ZLP, and thus enhances the detectability of these weak spectral features in the visible, mid- and far-infrared range. 

\subsection{Cathodoluminescence - CL}
\label{sec::methods::cl}
The excitations created by inelastic scattering in the sample (Fig. \ref{fig::Scattering_Sketch}) can propagate and eventually decay into other particles, bringing the sample back to its fundamental state. The emission of optical range photons (near infrared to far ultraviolet, NIR-UV) following electron excitation is known as cathodoluminescence (CL) \cite{Yacobi2013}. This well-known method of generating light has been used in electron microscopes since the early 1980s, with the observation of light emission from diamond \cite{Pennycook1980, Yamamoto1984}. The technique in STEMs received renewed interest following the seminal work of N. Yamamoto and collaborators on the light emission from small silver particles \cite{Yamamoto2001}.

As mentioned in Section \ref{sec::methods::eels}, electron excitation of materials is broadband. This is generally seen as a significant benefit for electron spectroscopies, as wide spectral ranges, some of which are hardly accessible with photon experiments, can be measured in single exposures. However, this implies that for a given emission energy on a CL spectrum, the excitation leading to it most likely occurred off-resonantly.

For this reason, in the current CL community, experiments can be divided into two categories: i) coherent CL and ii) incoherent CL. Broadly speaking, the first group includes experiments with surface plasmons \cite{Yamamoto2001, Coenen2011, Losquin2015, Kociak2017}, transition radiation \cite{Coenen2011} and photonic modes \cite{Auad2021}. For these, CL spectra can be directly compared to optical scattering \cite{Losquin2015}. The second group includes, in general terms, all experiments for materials with a bandgap, such as semiconductors \cite{Zagonel2011}, and possibly containing mid-gap states like point defects or color centers \cite{Tizei2012}. In these cases, CL spectra closely match those of off-resonance photoluminescence, as demonstrated by Z. Mahfoud and collaborators for group II-VI quantum dots \cite{Mahfoud2013}.

The historical model for CL states that most of the IR-UV photons emitted in CL emission stem from the decay of bulk plasmons \cite{Rothwarf1973,Meuret2015}. As bulk plasmons usually occur in the 20--30 eV range for semiconductors, each decay generates multiple electron-hole pairs at or above the bandgap energy of the material \cite{Rothwarf1973}, and multiple photons may be emitted as photon bunches \cite{Meuret2015}. Furthermore, the electron-hole pairs can have a larger energy, possibly explaining why some defects or color centers are observable in photoluminescence and not in CL, and vice-versa \cite{Zaitsev2013}, as is the case for the neutral and charged states of the nitrogen-vacancy center in nanodiamonds \cite{Tizei2012, Sola2019}.

\subsection{Electron energy gain spectroscopy - EEGS}
\label{sec::methods::eegs}

A final electron spectroscopy technique that probes excitations in the optical range is electron energy gain spectroscopy (EEGS) \cite{Abajo2008, Henke2021, Auad2022}. It is a lesser-known and less frequently used technique, but it has the potential to revolutionize electron spectroscopy for nano-optics. In this technique, a photon beam is shone onto the sample simultaneously with the electron beam. In the presence of the electromagnetic near-field of the sample, photons can be absorbed by the electron, leading to (stimulated) energy gain. This has been previously explored in photon-induced near-field electron microscopy (PINEM) \cite{Barwick2009}. The coupling probability depends on the intensity of the incident photon beam and on the density of electromagnetic modes at a given frequency \cite{Abajo2008}. By tuning the laser energy, one can map this coupling probability, which at constant photon intensity is proportional to the density of electromagnetic modes. The spectrum created in this fashion is named the EEGS spectrum. The energy resolution achieved in EEGS is given by the photon beam spectral line width, which can be orders of magnitude better than that achievable with electron beams even with aid from the latest generation of electron monochromators. This spectroscopy has been recently demonstrated with optical modes in ring-resonators \cite{Henke2021} and dielectric spheres \cite{Auad2022}, with the record for energy resolution of 2 $\mu$eV \cite{Henke2021}.

The typical exciton linewidth at cryogenic temperatures for TMD monolayers \cite{Cadiz2017,Raja2019} is below the ultimate spectral resolution of the current technology for EELS (around 1 meV), making EEGS experiments on these materials particularly promising. Also, along with stimulated energy gain events, stimulated energy loss can also occur, providing a way to create optical excitations in resonance and with a high probability, something that is impossible with a free electron beam. Therefore, EEGS is a potentially very interesting technique for exploring the exciton physics of TMD monolayers.

\subsection{Cathodoluminescence excitation spectroscopy - CLE}
\label{sec::methods::cle}

One of the biggest benefits of electron spectroscopies, the broadband excitation, becomes a limitation when attempting to understand the exact excitation and decay pathways in materials. To circumvent this difficulty, the energy lost by each electron can be probed, in conjunction with the detection of an emitted particle. This time-coincidence spectroscopy has been demonstrated in the past for secondary electrons \cite{Pijper1991}, Auger electrons \cite{Haak1984}, X-ray photons \cite{Kruit1984}, and visible photons \cite{Ahn1984}. Some of these time-coincidence spectroscopies have been recently revisited using modern electron detection technologies \cite{Jannis2019, Feist2022, Varkentina2022}. N. \citet{Varkentina2022} explored CL-EELS time-coincidence spectroscopy as a nanoscale analogue of photoluminescence excitation spectroscopy (PLE), and showed that the photon emission probability varies as a function of electron energy loss. This technique, called cathodoluminescence excitation spectroscopy (CLE), will allow for a better understanding of CL generation mechanism in TMD monolayers.

\subsection{TMD sample preparation for electron microscopy/spectroscopy}
\label{sec::methods::samples}
There are numerous ways to obtain atomically thin and even monolayer materials, ranging from mechanical exfoliation of bulk crystals to chemical vapor deposition growth \cite{Novoselov2016, Castellanos2022}. Moreover, there is also a myriad of techniques to manipulate these thin layers and construct van der Waals heterostructures \cite{Novoselov2016, Castellanos2022}. Samples for experiments in a TEM have to be electron transparent, which limits sample thicknesses for CL and imaging/diffraction to below $\sim$150 nm and for EELS to below $\sim$50 nm. The actual limit depends on the primary electron kinetic energy.

The objective of this section is to briefly describe successful methods that have been used to prepare TEM-compatible samples, in particular 2D material heterostructures. The requirements for TMD samples for electron spectroscopy are ultimately more stringent than for other electron microscopy techniques, such as atomically-resolved imaging and scanning electron diffraction (also known as 4D-STEM). The main issues for the latter are mitigation against carbon contamination and electron-beam-induced damage, where freestanding TMD layers will still generally suffice. An effective strategy for minimizing knock-on damage is to form heterostructures, either encapsulating in graphene or supporting with graphene \cite{Lehnert2017} or thin h-BN \cite{VanWinkle2023} on the exit and entrance surface, respectively. Since electron spectroscopy is accessing optical information, it requires samples of high optical quality with a controlled dielectric environment, which can only be achieved by fabricating in the form of vdW heterostructures due to criteria that will be discussed in later sections.

The first method for deterministic transfer of 2D materials is an all-dry viscoelastic polymer stamping with polydimethylsiloxane (PDMS) \cite{Castellanos-Gomez2014}, where individual layers are exfoliated directly onto PDMS and released onto the desired substrate one by one. Some disadvantages of this approach are: i) the PDMS comes into contact with all layer top surfaces, which can lead to poor sample cleanliness; ii) the repeated pressing requires more robust holey silicon nitride TEM supports that are prone to charging under the electron beam.

The second method is a van der Waals pick-up method, which is a variation of the first method involving an additional carrier polymer film that can increase the pick-up temperature \cite{Zomer2014,Pizzocchero2016}. Interfacial cleanliness is less compromised for heterostructure assembly with this method because the vdW interaction is used to pick up the subsequent 2D flakes exfoliated onto SiO$_2$/Si substrates. As such, only the top-most layer surface will come into contact with the polymer. In short, a thin polymer film (such as polycarbonate, PC, or polypropyl carbonate, PPC) on top of a PDMS stamp is used to sequentially attach the thin 2D materials required to form the complete van der Waals heterostructure. The thin materials are attached at moderately high temperatures (120$^\circ$C for PC), with the full heterostructure being built on the PC film/PDMS stamp assembly \cite{Shao2022}. The PC, with the attached heterostructure, is melted (at 180$^\circ$C) when in gentle contact with the amorphous carbon film of the TEM support grid. The residue of the PC film is then washed in chloroform solution \cite{Shao2022} or vapor.

Alternatively, the van der Waals heterostructure can be constructed on top of a Si/SiO$_2$ substrate following the two previously discussed methods. The final structure can be lifted using a wet transfer with a hydrophobic polymer film support like polymethylmethacrylate (PMMA) by etching the SiO$_2$ layer in a KOH solution. The heterostructure can then be placed on a TEM grid by dissolving the PMMA film using acetone, followed by cleaning steps in isopropyl alcohol (IPA) \cite{Bonnet2021}. The two methods are not all-dry as solvent washing of any polymer used is involved, however the heterostructure interfaces should see little contact with any polymer or solvent.

TEM grid support choice can be a matter of personal preference and transfer station heating stage setup, and each type has their own dis/advantages for the transfer methods described above. Most groups prefer the holey silicon nitride support films, where typically a 200 nm-thick amorphous SiN$_x$ membrane on a Si support frame has been patterned with micron-sized periodic holes, due to their robustness, chemical and heat resistance. Only layers suspended over the holes are electron transparent, however, and can complicate sample navigation with the added beam charging. The thick Si$_3$N$_4$ membrane, can sustain strong phonon-polariton and optical guided modes which can contribute significantly to losses in the IR-VIS ranges, complicating EELS experiments due to signal delocalization in this energy range. Metallic TEM grid support (commonly of Cu, Au, Mo, or Ni) with a tens of micron-sized square mesh overlaid with an additional thin 10--20 nm amorphous carbon support film with lacey webbing or regular array of micron-sized holes (such as Quantifoil or C-flat). This type of TEM grid is flexible for solution handling, electron transparent and conductive throughout for improved sample visualization, but the amorphous carbon support can also rupture easily.

\section{Electron energy-loss spectroscopy of suspended TMD monolayers}
\label{sec::init_eels}

EELS experiments in bulk TMDs date back to as early as the 1960s \cite{Wilson1969}. These electron scattering experiments were not performed in electron microscopes and, therefore, had very poor spatial resolution. For TMDs, to start with, one would need a spectral resolution better than $\sim$40 meV to be able to separate the two lowest energy excitons (X$_A$ and X$_B$ valley excitons) in Mo-based TMDs, which are split by spin-orbit coupling. The main limitation for high-resolution experiments was the available electron monochromator producing high-brightness beams, comparable to the brightness of cold field emitter electron sources. In the late 1990s, electron monochromators in electron microscopes creating electron beams with below 40 meV energy spread were already constructed \cite{Terauchi1999}. However, they were not stable enough to ensure this energy spread for long (10$^0$ s timescale) acquisitions, which is needed for the detection of excitonic transitions in TMD (as described in detail in Section \ref{sec::current_eels}).

\begin{figure} 
\centering{\includegraphics[width=0.9\textwidth]{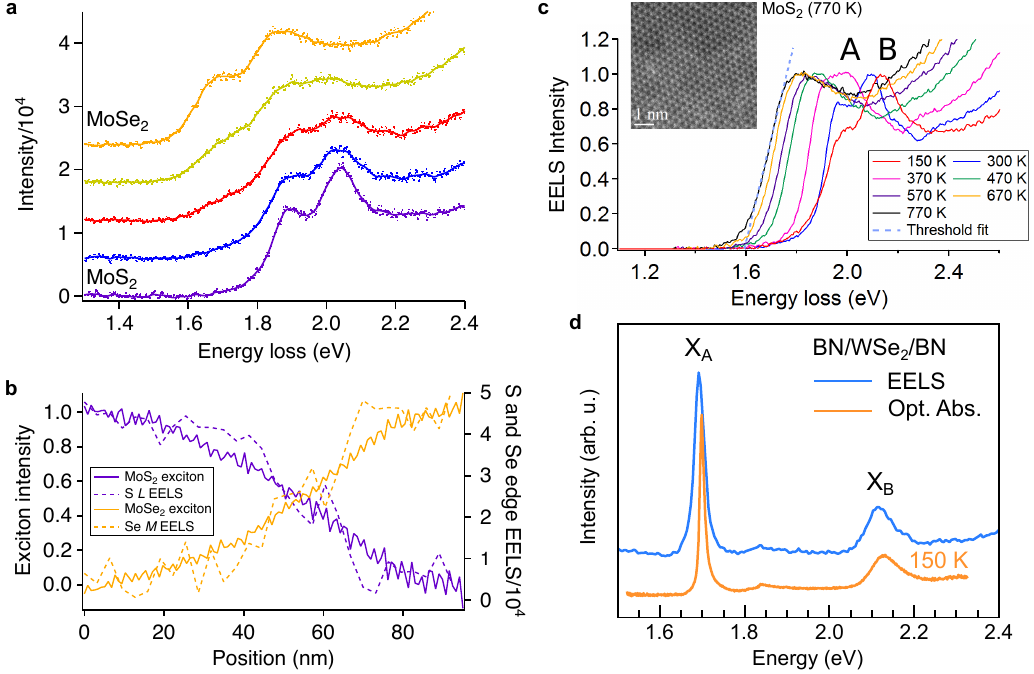}}
\caption{\textbf{a} Five spectra integrated at different positions across the interface a MoS$_2$-MoSe$_2$ chemically diffuse interface showing the excitonic A and B transitions. \textbf{b} Comparison of the fitting coefficient profiles with the chemical profiles measured from core-loss EELS of the S L and Se M edges on the same interface. \textbf{c} Temperature-dependent absorption spectra for MoS$_2$, where a shift towards lower energy in the onset (noted by the light-blue linear fitting curve) is observed with increasing temperature. \textbf{d} Comparison of EELS and optical absorption spectra of h-BN encapsulated WSe$_2$. Panels a,b adapted with permission from ref. \cite{Tizei2015}, panel c adapted with permission from ref. \cite{Tizei2016}, panel d adapted with permission from ref. \cite{Woo2024}.
\label{fig::EELS_FirstInterface}}
\end{figure}

In ref. \cite{Tizei2015}, the first EELS measurement in a STEM on TMD monolayers was capable of distinguishing the spin-orbit splitting between the two first excitonic transitions. In this report, the authors measured the spectral change as a function of position across a chemically diffuse MoSe$_2$/MoS$_2$ interface (Fig. \ref{fig::EELS_FirstInterface}(a--b)). Despite the chemical diffusion, no chemical shift of the excitonic transitions was observed, only an increase in linewidth, possibly due to the delocalization in electron scattering \cite{Muller1995, Egerton2011}. A clear limitation of this work is the linewidth of the excitonic transitions ($\approx$100--200 meV), even for liquid-nitrogen cooled (150 K) samples. Numerous studies have probed the nanoscale optical properties of TMDs using EELS since then, many of which were made at room temperature \cite{Dileep2016,Tinoco2017,Yan2020,Hong2021,Pelaez-Fernandez2021} reporting same range of linewidths. 

Temperature-dependent EELS experiments on MoSe$_2$ and MoS$_2$ monolayers were performed \cite{Tizei2016} from T = 150--770 K. The expected energy shift as a function of temperature \cite{Tizei2016} was reported, as shown in Fig. \ref{fig::EELS_FirstInterface}(c), in agreement with what was observed for TMD monolayers using optical spectroscopy \cite{Arora2020}. \citet{Gogoi2019} later examined WSe$_2$, MoS$_2$ monolayers and their heterostructures from T = 298--973 K, and similarly observed the expected redshift and peak broadening due to phonon scattering with temperature for all cases, including the intralayer excitons in the MoS$_2$/WSe$_2$ heterostructure. However, the linewidth broadening in both studies was much larger than expected, even at room temperature. In hindsight, this was probably due to surface contamination and electron beam damage in freestanding TMD monolayers. These large EELS linewidths were not understood until the work of N. \citet{Bonnet2021} and F. \citet{Shao2022}, in which hexagonal boron nitride (h-BN) encapsulation was shown to significantly sharpen exciton peaks in Figure \ref{fig::EELS_FirstInterface}(d). In these studies with the samples cryogenically cooled to T = 100--150 K, the linewidths were comparable (factor of 2 broader compared to factor of 10) to those observed by purely optical experiments in the same temperature range, as described subsequently in Section \ref{sec::current_eels}. With the improvement of optical quality, EELS spectra of TMDs monolayers can be directly compared to those of optical absorption \cite{Bonnet2021,Woo2024}, as shown in Figure \ref{fig::EELS_FirstInterface}(d) for h-BN encapsulated WSe$_2$.

As already discussed in Sec. \ref{sec::methods::eels}, EELS can excite a non-zero momentum transfer ($q \neq 0$) to measure the dispersive nature of TMD excitons, unlike optical techniques whose excitation has zero momentum transfer. Momentum-transfer or $q$-resolved EELS has shown that for WSe$_2$ monolayers, the A exciton dispersion is parabolic ($E = \hbar^2q^2/m_e$, with $m_e$ being the electron effective mass) \cite{Habenicht2015, Hong2020}. Additional absorption below the bandgap at finite momentum transfer ($q=0.1 \mathrm{\AA}^{-1}$) was detected in all four main types of semiconducting TMDs, namely WS$_2$, MoS$_2$, WSe$_2$ and MoSe$_2$, which the authors attributed to absorption states generated by chalcogen vacancies \cite{Hong2020}.

Higher-energy exciton resonances are generally not probed in optics, and their physical origin can be complex and not often discussed in literature. \citet{Hong2021} used the measured dispersion of these high-energy exciton resonances to decipher their complex physical origin difference between Mo- and W-based TMD monolayers and platinum dichalcogenides \cite{Hong2022} with the aid of Bethe-Salpeter equation (BSE) calculations.
This type of calculations showed that for TMD monolayers, the oscillator strength of the lowest-energy excitonic transition (X$_A$) is concentrated around the $K/K^\prime$ points, while higher energy transitions, including the spin-split X$_B$ transition has significant contributions away from $K$ \cite{Woo2023, Bergmann2024}.

\section{Electron energy-loss spectroscopy of suspended TMD homo- and hetero-bilayers}
\label{sec::bilayers}

The relative ease in generating van der Waals few-layer devices creates the opportunity to search for new and interesting physics in moir\'e structures \cite{Mak2022, Huang2022, Regan2022}. For example, in bilayer graphene at specific twist magic angles, superconductivity has been observed due to the formation of flat bands in their band structure \cite{Cao2018, Yankowitz2019}, or interface ferroelectricity with out-of-plane polarization in AB-stacked bilayer h-BN \cite{Yasuda2021} and twisted MoS$_2$ bilayer \cite{Weston2022} due to the reconstructed domains. EELS is a natural technique to explore modulations in the optical response of bilayer TMD homo- and heterostructures, given the nanometer-scale spatial resolution.

\begin{figure*} 
\centering{\includegraphics[width=0.9\textwidth]{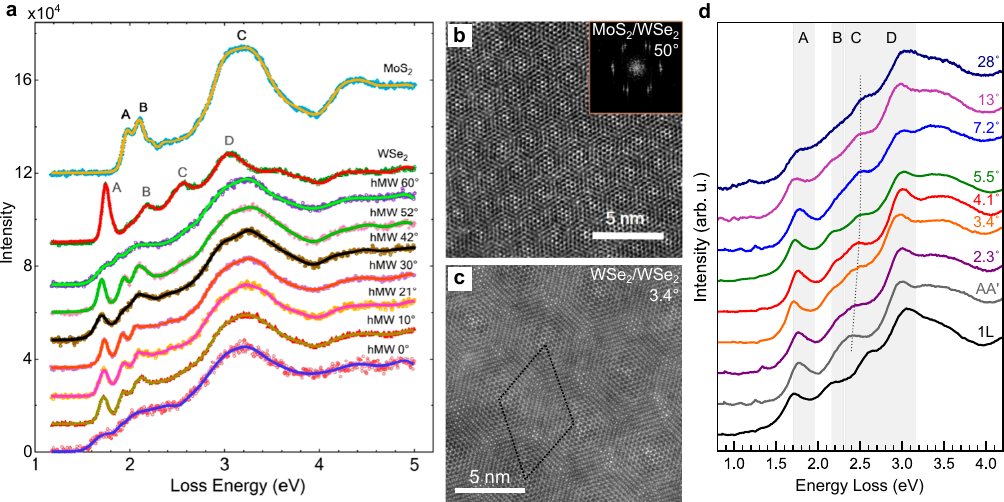}
\caption{Twist angle-dependent EELS spectra of MoS$_2$/WSe$_2$ van der Waals heterostructures and twisted bilayer WSe$_2$. \textbf{a} Comparison of the EELS spectra of monolayer MoS$_2$ and WSe$_2$ with that of aligned (anti-aligned) and misaligned van der Waals heterostructures. \textbf{b,c} STEM-HAADF images of 50$^\circ$ MoS$_2$/WSe$_2$ heterostructure and its inset image fast Fourier transfer (b), and WSe$_2$ bilayer with 3.4$^\circ$ relative twist angle (c). \textbf{d} EELS spectra from twisted bilayer WSe$_2$ with various twist angles compared to monolayer (1L) WSe$_2$ and AA$^\prime$ stacked bilayer. The dotted line highlights the change in the C exciton energy positions. Panels \textbf{a,b} adapted with permission from ref. \cite{Gogoi2019}, panels \textbf{c,d} adapted with permission from ref. \cite{Woo2023}.}
\label{fig::moire_bilayer}}
\end{figure*}

P. \citet{Gogoi2019} observed that in twisted MoS$_2$/WSe$_2$ heterostructures, the intralayer excitonic transitions have linewidths which strongly depended on the twist angle. For aligned (0$^\circ$ \textit{R}-type stacking) and anti-aligned (60$^\circ$ \textit{H}-type stacking) cases, the excitonic transition linewidths are significantly broader (Fig. \ref{fig::moire_bilayer}(a)) than in the respective free-standing monolayers or in bilayers with intermediary twist angles \cite{Gogoi2019}. Also, a systematic redshift of the WSe$_2$ A exciton occurs in bilayers, and towards multilayers (of WS$_2$ in this case \cite{Bergmann2024}). This layer-dependent energy shift is interpreted as a direct consequence of the increase in screening by the additional layer, as previously reported in optical studies \cite{Fang2014,Arora2015,Raja2017}. The large exciton linewidth broadening was attributed to the fast decay of intralayer excitons into direct interlayer excitons, which are favored when $K/K^\prime$ valleys in reciprocal space align between the two layers for 0$^\circ$ and 60$^\circ$ bilayers \cite{Gogoi2019}.

Considering homobilayers, ref. \cite{Woo2023} reported on the influence of twist angle in bilayer WSe$_2$ on the higher energy excitonic transitions, which involve electronic states whose chalcogen atom $p$-orbital character has out-of-plane components that are more sensitive to interlayer separation. The so-called \textit{C} excitonic transition increases in energy for larger twist angles in WSe$_2$, with a blueshift of up to 200 meV compared to the normal AA$^\prime$ stacking shown in Fig. \ref{fig::moire_bilayer}(d). To understand the origin of the blueshift, the authors considered how electronic states in reciprocal space contribute to this transition. The $GW+$BSE (Bethe-Salpeter equation) method on top of density functional theory (DFT) calculations showed that in WSe$_2$ bilayers, the C peak has contributions from multiple transitions at both $K$ and $Q$ points in reciprocal space. DFT calculations of WSe$_2$ bilayers over a range of twist angles demonstrated that the $K-Q$ gap increases in energy with increasing twist angle towards 30$^\circ$, explaining the trends observed for the C excitonic transitions \cite{Woo2023}.

S. \citet{Susarla2021} attempted to observe the spatial modulation of interlayer excitons by spatially resolved EELS in MoS$_2$/WSe$_2$ bilayer heterostructures. The broadened exciton peaks observed by \citet{Gogoi2019} in aligned bilayers ($\sim$0$^\circ$ \textit{R}-type stacking) were reproduced. A possible absorption peak linked to an interlayer excitonic transition, which varies in space within the moir\'e periodicity, was detected \cite{Susarla2021}. However, because of the weak oscillator strength of this transition and the broad linewidths intrinsic to free-standing bilayers, the signal could barely be detected.
To improve the possibility of detecting nanometer-scale spatial variations of excitonic transitions with EELS, S. \citet{Susarla2022} considered the modification of the A excitonic transition in WS$_2$ in a WS$_2$/WSe$_2$ bilayer heterostructure. The oscillator strength of the intralayer WS$_2$ A excitonic transition is much larger than that of moir\'e excitons \cite{Jin2019}. For this reason, nanometer scale variations of the A excitonic transition were detected between regions of different stacking in a moir\'e structure in a rotationally aligned WS$_2$/WSe$_2$ heterostructure \cite{Susarla2022}.

The next section will describe how h-BN, another 2D material of interest for optical applications, enabled the realization of the first CL experiments on TMD monolayers by \citet{Zheng2017}. This change in sample support also led to a dramatic decrease in absorption linewidth, as shown in Fig. \ref{fig::EELS_TMD_MonoAndEncapsulated}, which will be described in Section \ref{sec::current_eels}.

\section{Cathodoluminescence emission from TMD monolayer heterostructures}
\label{sec::cl}
\begin{figure} 
\centering
\includegraphics[width=0.5\textwidth]{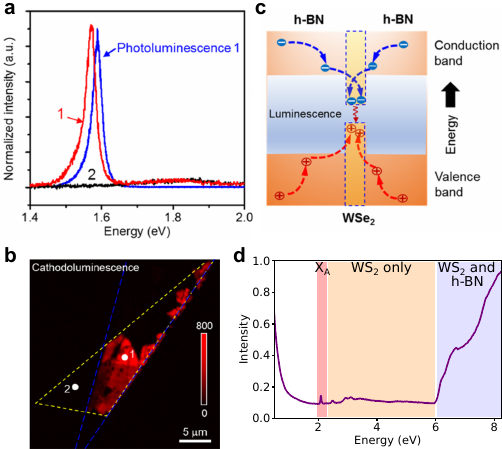}
\caption{\textbf{a} CL spectra at positions 1 and 2 in (b) are shown together with a PL spectrum acquired at position 1 in (b). \textbf{b} Monochromatic CL map of the h-BN/WSe$_2$/h-BN heterostructure filtered at 1.66 eV, the WSe$_2$ neutral exciton emission energy (color coded in red). The yellow outlines the WSe$_2$ monolayer, and the blue outlines the top h-BN layer. \textbf{c} Schematic showing the process of the generation, diffusion, and recombination of electron-hole (e-h) pairs. The minor number of e-h pairs generated in the WSe$_2$ layer is ignored in this model. Most electron inelastic scattering occurs in the h-BN layers \cite{Bonnet2021}, as shown in panel \textbf{d} of a low-loss EELS spectrum of h-BN encapsulated WS$_2$ monolayer. Panels a--c reproduced with permission from ref. \cite{Zheng2017}, panel d adapted with permission from ref. \cite{Bonnet2021}.}
\label{fig::CL_FirstbyZheng}
\end{figure}

The bright PL emission from TMD monolayers \cite{Mak2010} and the rich physics available due to their spin-valley states \cite{Xiao2012} attracted a lot of interest in this material, including those conducting CL in SEMs and STEMs in the early 2010s. Around this period, experiments were attempted in different laboratories with CL capabilities, but without any success. The major reasons for this were due to poor sample quality, excitation efficiency, and beam damage minimization. The pioneering work of S. \citet{Zheng2017} demonstrated that h-BN encapsulation made CL emission from a WSe$_2$ monolayer possible (Fig. \ref{fig::CL_FirstbyZheng}(a)). In the same work carried out in a SEM operating at 5 keV, the CL emission of WS$_2$ and MoS$_2$ monolayer samples encapsulated in the same fashion was also reported. By comparing the PL emission from the same sample, S. Zheng \textit{et al.} attributed the CL emission to neutral excitons (X$_A^0$). Moreover, with the sample cooled to 10 K, the charged exciton (trion, X$^-$) peak was detected at lower energy, indicating indeed that the first emission is due to the neutral exciton \cite{Zheng2017}. The CL emission intensity was mapped across the sample, showing a strong spatial variation due to cracks in the monolayer and residue bubbles (Fig. \ref{fig::CL_FirstbyZheng}(b)). According to their proposed model (Fig. \ref{fig::CL_FirstbyZheng}(c)), the h-BN encapsulation increased the TMD monolayer excitation probability, as excitations created in the larger band gap of h-BN could diffuse and recombine in the monolayer. This model implies that the CL emission intensity should increase with thicker h-BN layers, which was confirmed experimentally \cite{Zheng2017}. On the other hand, thicker encapsulation layers can decrease the spatial resolution in the case of SEM-CL \cite{Francaviglia2022}, which is delimited by the carrier diffusion length in h-BN.

\begin{figure*} 
\centering{\includegraphics[width=0.5\textwidth]{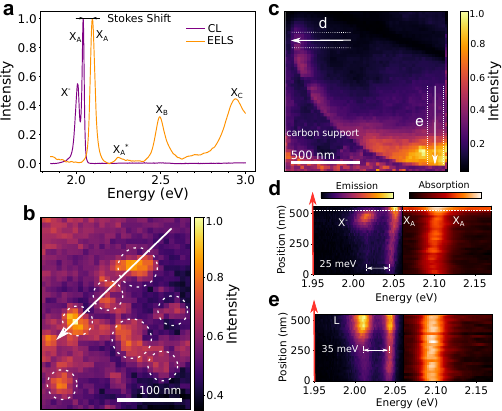}}
\caption{\textbf{a} EELS (orange) and CL (purple) spectrum from h-BN encapsulated WS$_2$ monolayer. The X$_A$, X$_B$, X$_C$, and X$^-$ (trion) peaks are labelled. The extra absorption between X$_A$ and X$_B$ is attributed to the $2s$ excited state of X$_A$, marked X$^*_A$. The curve intensities are normalized to match the X$_A$ maxima. \textbf{b} X$^-$ intensity map. The intensity was normalized by the maximum of the trion emission. High-intensity spots were marked by dashed circles. \textbf{c}  X$^-$ intensity map, normalized by the maximum of trion emission. \textbf{d--e} CL (left) and EELS (right) spectra along each arrow marked in (b) with spatial width of 100 nm. Figure panels modified with permission from ref. \cite{Bonnet2021}.}
\label{fig::CL_EELS_encapsulated}
\end{figure*}

Most electron energy-loss events do not take place in the TMD monolayer but in the thicker h-BN encapsulation layers, as shown for an h-BN/WS$_2$/h-BN heterostructure (Fig. \ref{fig::CL_EELS_encapsulated}(d) \cite{Bonnet2021}). Clearly, full h-BN encapsulation is a necessary condition, as all subsequent TMD monolayer CL experiments in both SEM and STEM to date \cite{Nayak2019,Singh2020,Bonnet2021,Francaviglia2022,Fiedler2023,Ramsen2023,Woo2024,Bonnet2024,Borghi2024} used h-BN encapsulation. In addition, later experiments showed that interface cleaning \cite{Nayak2019, Bonnet2021, Shao2022} is a crucial point to facilitate efficient charge carrier transfer, which is induced by residue expulsion into bubbles, a point already known to the TMD optics community \cite{Rosenberger2018,Purdie2018}. The tens of nm-thick h-BN encapsulation layer also provided remarkable protection against electron beam-induced damage for experiments at 60 keV and even up to 100 keV \cite{Bonnet2021}.

CL emission from TMD monolayers is dominated by the neutral exciton emission \cite{Zheng2017}. This emission line can be distinguished from the charged exciton line (Figure \ref{fig::CL_EELS_encapsulated}(a)) at sufficiently low temperatures when $kT$ ($k$ the Boltzmann constant and $T$ the temperature) is less than the trion binding energy (few tens of meV) \cite{Zheng2017, Bonnet2021, Woo2024}. With the sample cooled by liquid nitrogen, spatial variations of the trion (X$^-$) emission down to the tens of nanometer scale were observable with the improved spatial resolution in (aberration-corrected) STEM as shown in Figure \ref{fig::CL_EELS_encapsulated}(b) \cite{Bonnet2021}. The authors associated these variations to changes in the local dielectric environment in the presence of carbonaceous residues from the sample preparation.

An open question remains regarding the direct comparison of TMD CL and off-resonance photoluminescence (PL) spectra. Up to the time of writing, only \citet{Zheng2017}, \citet{Francaviglia2022}, and \citet{Fiedler2023} have reported combined CL and PL measurements of the same sample. All three studies reported a redshift of the CL emission with respect to the PL emission, which the authors have associated to local heating by the electron beam or slight differences in the measurement area.

A direct consequence of the first CL experiments in a monochromated STEM also capable of high spectral resolution EELS ($<$ 20 meV) was the observation that the exciton linewidths in EELS are significantly narrower (tens meV vs. hundreds meV) in h-BN encapsulated TMD monolayers (Figure \ref{fig::CL_EELS_encapsulated}(a)). The reasons and significance of this observation are discussed in the next section. Finally, combined EELS and CL allow for the spatial mapping of the Stokes shift (Figure \ref{fig::CL_EELS_encapsulated}(c--e)), which can elucidate exciton-phonon coupling effects that dominate towards higher temperatures. It should be noted however that this is partially hindered by the EELS energy scale calibration accuracy (of the order of 1 \%, so 20 meV at 2 eV).

\section{Improvement of exciton linewidths by encapsulation}
\label{sec::current_eels}

\begin{figure} 
\centering{\includegraphics[width=0.9\textwidth]{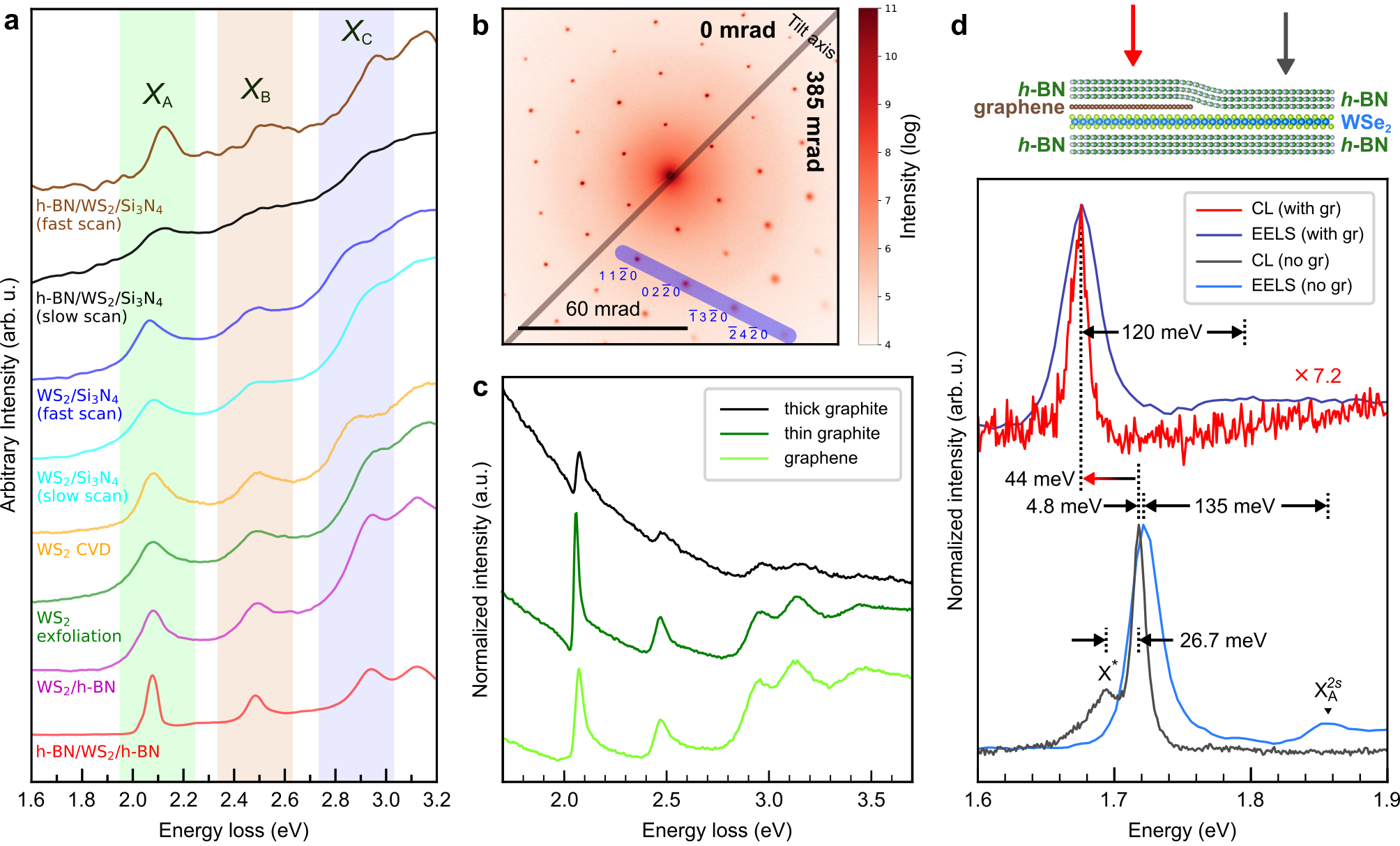}}
\caption{\textbf{a} EELS spectra of WS$_2$ monolayers on different substrate configurations at 110 K: h-BN encapsulated (red), h-BN supported (purple), freestanding exfoliated (green), freestanding CVD-grown (yellow), slow-scan Si$_3$N$_4$ supported (cyan), fast-scan Si$_3$N$_4$ supported (blue), slow-scan Si$_3$N$_4$/h-BN encapsulated (black), and fast-scan Si$_3$N$_4$/h-BN encapsulated (brown) cases. Each spectrum is normalized to its X$_A$ intensity after zero-loss peak alignment and tail subtraction. \textbf{b} A composite electron diffraction pattern of a freestanding exfoliated WS$_2$ monolayer with no sample tilt (upper left), and high sample tilt (lower right, 385 mrad or $\sim$22$^{\circ}$) where the spots have become diffuse. \textbf{c} EELS spectra of WS$_2$ monolayer encapsulated between two graphite layers of total thickness of 17.5 nm (thick), 6.1 nm (thin), and two graphene layers (0.69 nm), each showing the asymmetric lineshape. \textbf{d} CL and EELS spectra from identical regions with the emission peaks assigned, including a prominent low-energy trion (X$^*$) emission in the absence of graphene (lower), as well as a clear quenching of the neutral exciton ($\sim$7$\times$ lower) but no trion emission with graphene (upper). The change in energy separation between the $1s$ and $2s$ Rydberg states are also highlighted. Panels \textbf{a,b} modified with permission from ref. \cite{Shao2022}, panels \textbf{c,d} modified with permission from ref. \cite{Woo2024}.
\label{fig::EELS_TMD_MonoAndEncapsulated}}
\end{figure}

The observation of excitonic transitions with significantly narrower linewidths in EELS for TMD monolayers encapsulated in h-BN \cite{Bonnet2021} motivated a study by F. \citet{Shao2022} to understand the role of h-BN in the improved optical response (Fig. \ref{fig::EELS_TMD_MonoAndEncapsulated}(a--b)). The authors considered three characteristics of h-BN that could improve the optical response of the TMD monolayer proposed by the optical community \cite{Cadiz2017,Ajayi2017,Wierzbowski2017}: i) decreased roughness of monolayers; ii) improved cleanliness; and iii) reduced substrate-induced charge accumulation. Few techniques can address the various proposed factors single-handedly to comprehensively compare and disentangle their contributions to inhomogeneous linewidth broadening. Electron spectro-microscopy, combining monochromated EELS for the excitonic response, elemental identification of impurities with core-loss EELS, and electron diffraction tilt series, was able to overcome such a challenge to rank the dominant effects on linewidth.

For graphene, electron diffraction experiments have shown that when supported or encapsulated by h-BN, it could significantly reduce its atomic roughness 5- and 10-fold, respectively \cite{Meyer2007, Thomsen2017}. The same is true when the TMD monolayers are supported in h-BN: their roughness is a factor of 5 smaller than their free-standing counterparts \cite{Shao2022}. However, a rougher support, such as amorphous Si$_3$N$_4$, leads to a monolayer with larger roughness than free-standing monolayers with their intrinsic corrugation \cite{Shao2022}. Another atomically flat substrate, graphite, also leads to a reduction in TMD layer roughness when used for encapsulation \cite{Woo2024}.

Residue at any interface and free surfaces, inherent to the TMD exfoliation and transfer processes involved during sample preparation, also significantly impacts linewidths, as observed in purely optical experiments. However, because of the strong vdW forces between 2D materials in encapsulated structures and the gathering of interfacial residues into bubbles, regions free of contaminants can also be found adjacent to said bubbles.
F. \citet{Shao2022} also observed that specifically for amorphous Si$_3$N$_4$ substrates, the linewidth of the excitonic transitions depended on how the heterostructure was exposed to a given electron dose. The effect was reversible, indicating it did not occur due to damage generated by the electron beam. They interpreted this effect as charge trapping on the Si$_3$N$_4$ amorphous support that is prone to have dangling bonds, showing yet another advantage of thin h-BN flakes with low defect density.

In exploring the effects of other changes to the dielectric environment, encapsulation in conductive substrates like graphite and graphene has resulted in Fano-like line profiles in their absorption spectra \cite{Woo2024}, in contrast to almost Lorentzian lineshapes in free-standing and h-BN-encapsulated TMD monolayers (Fig. \ref{fig::EELS_TMD_MonoAndEncapsulated}(c--d)). This occurs due to how the electron beam couples to the thin metal/semiconductor heterostructures, as interpreted with a 2D optical conductivity model \cite{Woo2024}. This shows how interfacing TMD monolayers with graphene can lead to the dielectric engineering of their excitonic response, including the lineshape, exciton charge state (quenching luminescence from the charged exciton) \cite{Lorchat2020,Woo2024}, modifying the band gap as well as the exciton binding energy from changing Coulomb interaction strength \cite{Raja2017,Tebbe2023,Woo2024} (Fig. \ref{fig::EELS_TMD_MonoAndEncapsulated}(d)). The next section will discuss how coupling TMD excitons with different structures can exploit this further, and lead to a significant modification of their properties.

\section{Electron spectroscopies of TMDs coupled to plasmonic and photonic structures}
\label{sec::TMDcoupling}
Coupling of excitations can lead to unique physical phenomena. For example, weak coupling of excitations in an optical cavity modifies their decay rate (Purcell effect \cite{Anger2006}). Precise tuning of the excitonic response can be achieved by strong coupling in optical cavities \cite{Weisbuch1992,Lidzey1998}. Coupling to near fields, for example in plasmonic nanostructures, occurs at subwavelength scales, which necessitates the use of EELS and CL to study these interactions.

\begin{figure*} 
\centering{\includegraphics[width=0.7\textwidth]{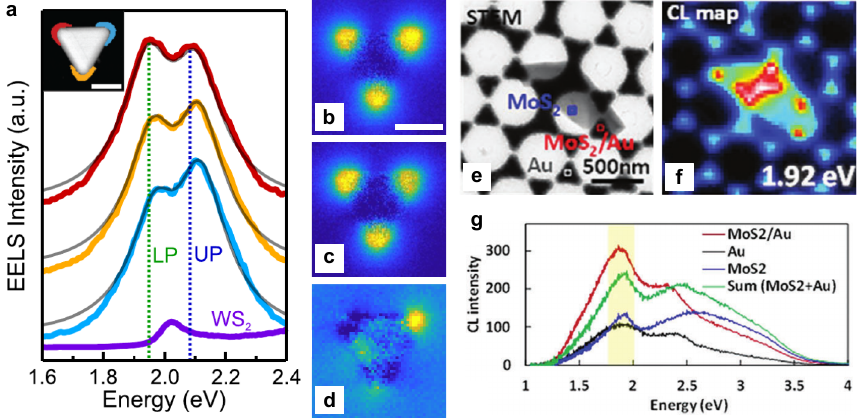}}
\caption{\textbf{a} EELS spectra from each corner of the coupled triangular Au nanoparticle on top of six layers of WS$_2$ shown in the HAADF STEM image in the inset. The lower (LP) and upper (UP) polariton peaks in the red spectra are marked by the vertical dotted lines at 1.94 and 2.07 eV, respectively. The gray lines are the fits of the experimental EELS spectra using the coupled mode theory. The plasmon-exciton coupling strengths, g, extracted through fitting, are 74 $\pm$ 10, 70 $\pm$ 11, and 67 $\pm$ 14 meV for the red, yellow, and blue spectra, respectively. The plasmon-exciton detuning energies ($\omega_{pl} - \omega_X$), extracted through fitting, are -6 $\pm$ 3, 14 $\pm$ 3, and 29 $\pm$ 5 meV for the red, yellow, and blue spectra, respectively. The purple EELS spectrum is from the uncoupled WS$_2$ flake far away from the metallic nanoparticle.  \textbf{b--c} EELS maps extracted at the UP and LP energies, respectively, from the spectrum image that has been normalized to the zero-loss peak height. \textbf{d} The difference map shows nanoscale variation in the coupling. The scale bar in (a,b) is 50 nm. \textbf{e--f} STEM bright-field image and CL map at 1.92 eV of a MoS$_2$ flake on top of a Au nanopyramid array. \textbf{g} CL spectra on pure MoS$_2$ (blue), the  Au nanopyramid array (black), and the MoS$_2$ on a nanopyramid (red). Panels \textbf{a--d} modified with permission from ref. \cite{Yankovich2019} and panels \textbf{e--g} modified with permission from ref. \cite{Vu2023}.
\label{fig::EELS_StrongCoupling}}
\end{figure*}

A. \citet{Yankovich2019} have studied the strong coupling of excitons in thin WS$_2$ to plasmons from Ag triangles. For a 6-layer thick WS$_2$ flake coupled to $\sim$50--60 nm wide Ag triangles, a splitting of $\sim$70 meV was achieved, which is sufficiently large to show mode hybridization given the exciton and plasmon linewidths of $\sim$70 and $\sim$210 meV, respectively. The coupled modes were mapped with nanometer-sampling (Fig. \ref{fig::EELS_StrongCoupling}(a--d)), showing spatial variations between the upper polariton (UP) and lower polariton (LP) modes that indicate where the system is positively and negatively detuned \cite{Yankovich2019}. A similar system, MoS$_2$ layers coupled to $\sim$10 nm wide Au islands, was studied by K. \citet{Reidy2023}. They observed a new peak some 80 meV below the MoS$_2$ A exciton, which they attributed to screening of the Coulomb interaction due to the proximity to the Au.

Thin flakes of TMDs are also optical cavities which can sustain optical modes that couple to light and that can be excited by fast electrons \cite{Taleb2022,Chahshouri2022,Vu2022,Davoodi2022}. The precise energy of these modes and how they couple to photons and electrons depend on the thin flake geometry and can be probed using CL hyperspectral mapping. Furthermore, coupling of TMD flakes to metallic structures has demonstrated CL enhancement of the neutral exciton emission, as shown by \citet{Vu2023} for MoS$_2$ on top of Au pyramids arrays in Fig. \ref{fig::EELS_StrongCoupling}(e--f). 

\section{Summary and Perspectives}
\label{sec::perspectives}
The suite of complementary characterization techniques available in the scanning and scanning transmission electron microscope makes electron spectroscopies like EELS and CL well-positioned to directly correlate the nano- and atomic-scale structural and chemical properties to the nanometric optical response in 2D materials. As such, the study of TMD exciton physics with EELS and CL is becoming more common, with an increase in laboratories with SEM and STEMs equipped with light injector/collectors for CL and electron monochromators for high spectral resolution EELS. Since the first STEM-EELS experiment on monolayer TMDs, significant progress has been made once the local dielectric environment could also be well-controlled in TEM-compatible specimens. In the form of vdW heterostructures, the measured EELS excitonic absorption linewidth approaches the limits of thermal broadening at liquid nitrogen temperatures ($\sim$100 K). Encapsulation with h-BN has concurrently enabled sufficient excitation into the TMD monolayer for CL emission. These endeavors have put EELS absorption and CL emission on par with optical spectroscopy techniques in the same temperature range, such that fine energy shifts between different exciton species or energy splitting of exciton states in the presence of a moir\'e potential are now measurable in the electron microscope. Furthermore, with the advantage of being below the diffraction limit, understanding the nature of TMD excitons at the (sub-)nanoscale is now a reality, as exemplified by the works highlighted in this review. Many limitations and technological challenges still exist, however.

The current record for EELS spectral resolution, coupled with high spatial resolution experiments of approximately 3 meV \cite{Lagos2022}, is still far larger than the intrinsic exciton linewidths for the best TMD monolayers (approximately 1 meV at 4--5 K \cite{Ajayi2017,Cadiz2017,Raja2019}). At present, generating 1 nm wide electron beams with a sub-1 meV energy spread seems improbable unless a radically new technology arises. EEGS is a possible alternative, as it can easily reach sub-meV spectral resolution. However, at the time of writing, EEGS of semiconductors has not yet been demonstrated.

An intrinsic limitation to EELS and CL experiments is the physics behind electron scattering in matter: the energy loss at each scattering event cannot be controlled, and only its probability can be calculated. This limitation can be overcome by temporal coincidence spectroscopy, as recently demonstrated \cite{Varkentina2022, Feist2022}. This technique, called CLE, allows for the measurement of the quantum efficiency of materials for light emission, similar to photoluminescence excitation spectroscopy (PLE) \cite{Varkentina2022}, and of excitation decay times \cite{Varkentina2023}. CLE could help identify the physical mechanisms behind light emission in h-BN/TMD/h-BN heterostructures excited by fast electrons. The observation of STML with the TMD monolayer outside the tunnel junction indicates that the mechanisms driving the excitation of TMDs with a tunnel current or free electrons through h-BN are not fully understood \cite{Wang2023}.

Concerning the single-photon sources in TMD monolayers \cite{Tonndorf2015, Parto2021}, two advancements could be envisaged. First, the preparation of 2D material heterostructures compatible both with high spatial resolution optical measurements (confocal PL or super-resolved optical techniques) and with high spatial resolution imaging in a TEM/STEM is a possible way to identify the atomic structure of defects. A second possibility would be to observe a stable single-photon source in TMDs using a Hanbury-Brown and Twiss interferometer coupled to a STEM-CL system, as done in other materials \cite{Tizei2013, Bourrellier2016, Fiedler2023b}. The main experimental obstacle at this time, is the availability of cooling to temperatures below 100 K, as necessary for most single-photon sources. In addition, CL might be a more adapted technique to measure confined excitonic states in moir\'e heterostructures (as very recently demonstrated by \citet{Borghi2024}), in which the oscillator strength of interlayer excitons in absorption is small \cite{Barre2022}. However, given the small confining potentials, cooling to below 50 K might be required. The situation could change in the near future, with the availability of STEM-CL-compatible holders with liquid helium cooling. Furthermore, compatibility with optoelectronic devices, such as to enable gate tunability or application of external fields, will also be beneficial in accessing dynamic modulation of doping concentration into the TMD. This is more straightforward to achieve within an SEM where sample thickness and substrate requirements are less demanding. It is anticipated that when these major technological constraints can be overcome in the transmission electron microscope, electron spectroscopies will play an even more prominent role in new discoveries towards the atomic-scale in TMDs and other 2D materials.

\section{Acknowledgements}

This work was supported by the Center for Nanophase Materials Sciences (CNMS), which is a U.S. Department of Energy, Office of Science User Facility at Oak Ridge National Laboratory. The authors acknowledge funding from the ANR JCJC Grant SpinE (reference no. ANR-20-CE42-0020), and funding from the European Union’s Horizon 2020 research and innovation programme under grant agreement No. 101017720 (EBEAM).

\bibliography{TMD_NanoOpticsWithElectrons.bib}

\end{document}